\documentclass[%
reprint,
superscriptaddress,
amsmath,amssymb,
aps,
floatfix,
]{revtex4-1}

\usepackage[utf8]{inputenc}
\usepackage{graphicx}
\usepackage{dcolumn}
\usepackage{bm}
\usepackage{siunitx}
\usepackage{balance}
\usepackage{xr}

\usepackage{xr}
\makeatletter
\newcommand*{\addFileDependency}[1]{
  \typeout{(#1)}
  \@addtofilelist{#1}
  \IfFileExists{#1}{}{\typeout{No file #1.}}
}
\makeatother

\newcommand*{\myexternaldocument}[1]{
    \externaldocument{#1}
    \addFileDependency{#1.tex}
    \addFileDependency{#1.aux}
}

\myexternaldocument{supp_qvhe}

\begin{document}

\preprint{}

\author{S. Arora}\thanks{These authors contributed equally to this work}
\affiliation{Kavli Institute of Nanoscience, Delft University of Technology, 2600 GA, Delft, The Netherlands}
\author{T. Bauer}\thanks{These authors contributed equally to this work}
\affiliation{Kavli Institute of Nanoscience, Delft University of Technology, 2600 GA, Delft, The Netherlands}
\author{R. Barczyk}\thanks{These authors contributed equally to this work}
\affiliation{Center for Nanophotonics, AMOLF, Science Park 104, 1098 XG Amsterdam, The Netherlands}
\author{E. Verhagen}
\affiliation{Center for Nanophotonics, AMOLF, Science Park 104, 1098 XG Amsterdam, The Netherlands}
\author{L. Kuipers}\email{l.kuipers@tudelft.nl}
\affiliation{Kavli Institute of Nanoscience, Delft University of Technology, 2600 GA, Delft, The Netherlands}

\title{Direct quantification of topological protection in symmetry-protected photonic edge states at telecom wavelengths}

\date{\today}

\begin{abstract}
Topological on-chip photonics based on tailored photonic crystals (PhC) that emulate quantum valley Hall effects has recently gained widespread interest due to its promise of robust unidirectional transport of classical and quantum information. We present a direct quantitative evaluation of topological photonic edge eigenstates and their transport properties in the telecom wavelength range using phase-resolved near-field optical microscopy. Experimentally visualizing the detailed sub-wavelength structure of these modes propagating along the interface between two topologically non-trivial mirror-symmetric lattices allows us to map their dispersion relation and differentiate between the contributions of several higher-order Bloch harmonics. Selective probing of forward and backward propagating modes as defined by their phase velocities enables a direct quantification of topological robustness. Studying near-field propagation in controlled defects allows to extract upper limits to topological protection in on-chip photonic systems in comparison to conventional PhC waveguides. We find that protected edge states are two orders of magnitude more robust than the latter. This direct experimental quantification of topological robustness comprises a crucial step towards the application of topologically protected guiding in integrated photonics, allowing for unprecedented error-free photonic quantum network

\end{abstract}
\maketitle

The emergence of photonic topological insulators (PTIs) has led to promising theoretical and experimental approaches for topology-protected light-matter interaction \cite{Klembt2018} and integration of robust quantum devices \cite{Carusotto2013}. Topologically protected photonic edge states offer robust energy transport with unprecedented guiding capabilities, providing a cornerstone for efficient distribution of classical and quantum information in dense networks \cite{Hasan2010,Soljacic2018}. The notion of such states supporting unhindered transport around defects and sharp corners is especially interesting for on-chip applications. In addition to Chern-type PTIs that break time-reversal symmetry \cite{Haldane2008, Raghu2008, Soljacic2018, Plotnik2014}, a time-reversal invariant realization of lossless optical transport was introduced theoretically on a dielectric photonic crystal (PhC) platform at telecom frequencies \cite{Wu2015, Khanikaev2013}. While the existence of these states has been evidenced in the linear \cite{Barik2018,Parappurath2020} and nonlinear regime \cite{Smirnova2019}, and topological robustness has been inferred by high transmission \cite{Shalaev2018, He2019}, quantifying their defining quality of scattering-free propagation has remained elusive. Potential interference effects and out-of-plane scattering losses at local disorder render such a quantification challenging. Here, we report a rigorous robustness evaluation of valley photonic edge eigenstates at telecom wavelengths. Local investigation of the states' transport properties via phase-resolving near-field microscopy gives direct insight into topological protection, through distinction between forward and backward waves. We find that protected edge states are two orders of magnitude more robust as compared to conventional waveguides. This determination of significantly suppressed back-reflection provides a crucial step towards implementing topological guiding in on-chip photonic networks.

We realize valley-Hall PhCs (VPCs), which rely on the valley degree of freedom linked to the breaking of a specific lattice symmetry \cite{Shalaev2018, Lu2014a, Lu2018, Kim2014,Dubrovkin2020}. Similar to the valley-selective polarization caused by spin-orbit coupling in transition metal dichalcogenides \cite{Dong2017}, these PhC lattices exhibit a non-vanishing Berry curvature at the $K$ and $K^{\prime}$ points of the Brillouin zone \cite{Tzuhsuan2016}.  Since each valley is associated with an intrinsic magnetic moment, the valley-Chern invariant $C_{K,K^{\prime}} = \pm 1/2$ signifies a \textit{pseudo-spin} \cite{Xiao2007}, rendering the bulk band structure topologically non-trivial. A domain wall formed by two parity inverted copies of the PhC lattice results in two degenerate and robust edge-state eigenmodes confined to the interface that linearly traverse the photonic band gap (PBG), each with a unique \textit{pseudo-spin} \cite{Zak1989}. As long as the lattice symmetry is preserved and no inter-valley scattering occurs to flip the \textit{pseudo-spin}, these edge states are predicted to be immune to reflection from local disorder along the domain wall \cite{Tzuhsuan2016,Cheng2016}.

\begin{figure*}[hbt]
  \centering
  \includegraphics[width=\textwidth]{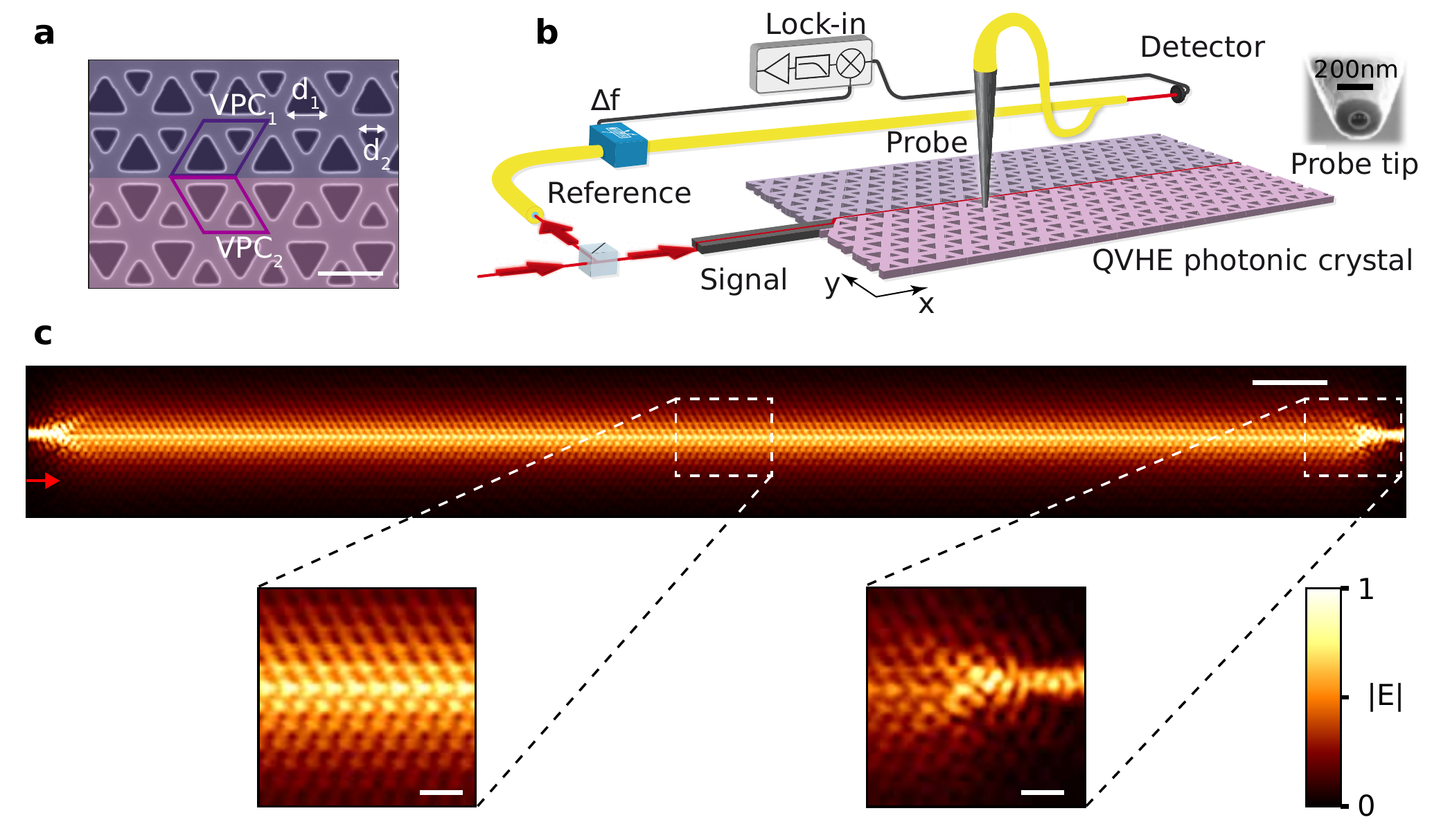}
  \caption{\textbf{Experimental visualization of a topological edge state in a valley photonic crystal. a,} SEM of the fabricated structure with two pseudo-colored regions depicting the two lattices VPC$_1$ and VPC$_2$, with opposite valley-Chern invariants. The unit cell with lattice constant $a = \SI{503}{nm}$ consists of equilateral triangular holes of side lengths $d_1 = 0.7a$ and $d_2 = 0.45a$. Scale bar: \SI{500}{nm}. 
  \textbf{b,} Schematic of the near-field scanning optical microscope used to map the in-plane field distribution of the topologically non-trivial PhC edge mode. To facilitate heterodyne-based phase detection, the input beam is split into two branches, labeled signal and reference. An aperture-based near-field probe collects part of the evanescent tail of the in-plane field components while scanning over the crystal at a height of $\SI{20}{nm}$, and couples the collected light to a single mode optical fibre. Inset: SEM image of the probe.  
  \textbf{c,} Measured  normalized amplitude of the in-plane field components at a laser excitation wavelength of $\lambda = \SI{1600}{nm}$  over the extent of 165 unit cells, with the scale bar corresponding to $\SI{5}{\mu m}$. Light is launched from a feed waveguide at the left side of the crystal, with the direction indicated by the red arrow. Left inset: Zoom-in of the detected field amplitude pattern along the domain wall. Right inset: Zoom-in of the out-coupling flank of the access waveguide. Scale bar: $\SI{1}{\mu m}$.  }
  \label{fig:1}
\end{figure*}

\begin{figure*}[hbt]
  \centering
  \includegraphics[width=\textwidth]{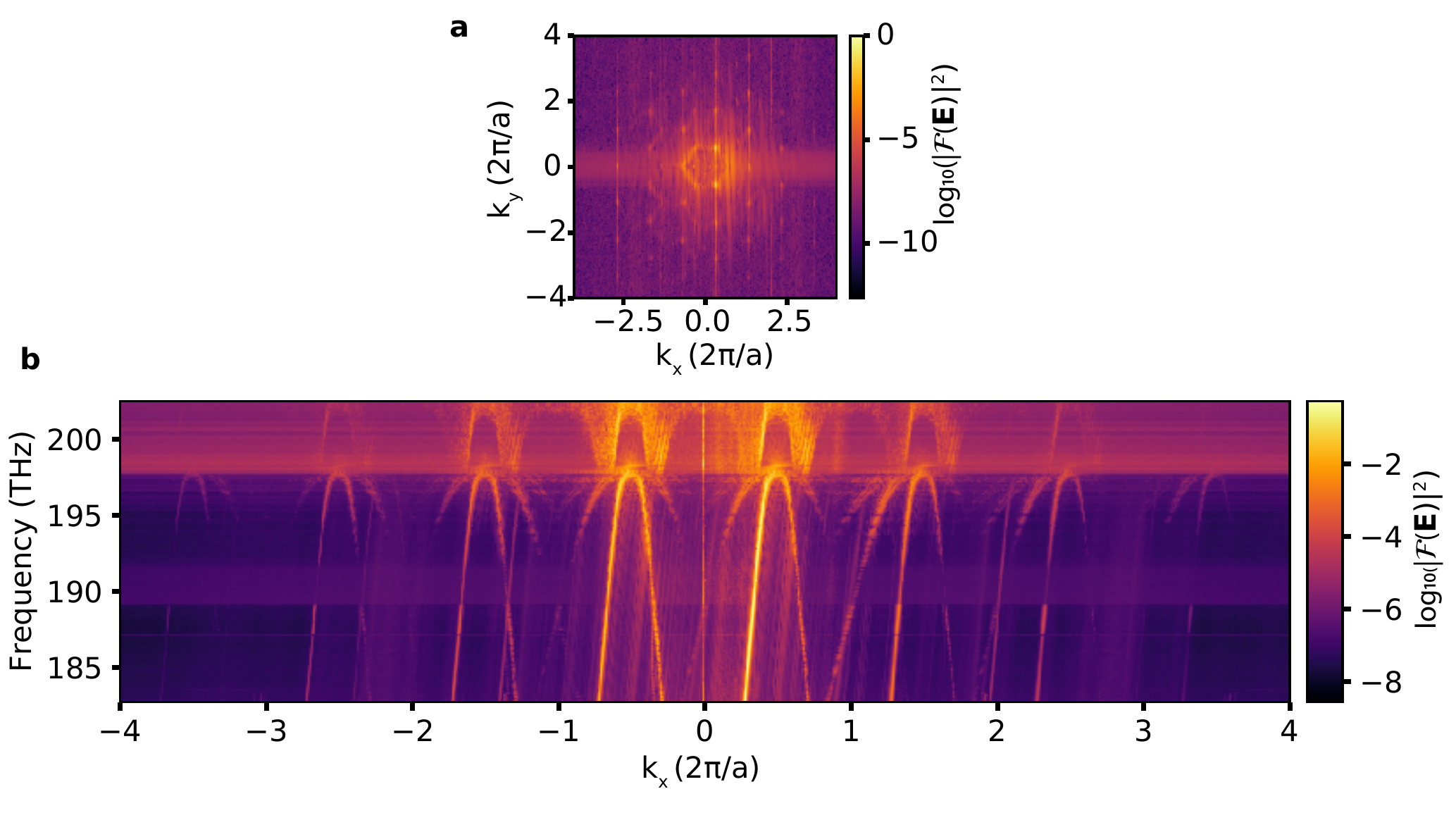}
  \caption{\textbf{Momentum space of the VPC edge state.} \textbf{a,} Two-dimensional Fourier transform of the real space amplitude distribution of the PhC mode. High intensity points are periodically separated by the reciprocal lattice vector $2\pi/a$ in the direction of propagation $k_x$ along the edge and by $4\pi/ \sqrt{3}a$ in the transverse direction $k_y$, representing the bulk reciprocal symmetry.
 \textbf{b,} Experimentally retrieved dispersion diagram. Bright lines of positive slope indicate positive group velocity (forward propagating modes), while lines with negative slope indicate a negative group velocity (backward propagating modes). Consecutive Bloch harmonics are separated by the size of a single Brillouin zone ($2\pi/a$). Frequencies above \SI{197.5}{THz} correspond to bulk bands. The Fourier intensity for each excitation frequency is normalized to the overall maximum value. In addition to the dominant modes in forward and backward direction, lines with half and a third of the dispersion slope appear. These are attributed to a nonlinear interaction with the scanning near-field probe (see Supplementary Sec. 3)}
  \label{fig:2}
\end{figure*}

To determine the experimentally achievable robustness against back-scattering, we fabricate a VPC working at telecom wavelengths on a silicon-on-insulator (SOI) platform following the design of \cite{Shalaev2018} (see Fig. \ref{fig:1}a). Light is coupled into the PhC structure in the +$x$ direction from an access waveguide. This system supports edge modes of opposite group velocity $\pm v_g$ (see Supplementary Fig. S1) propagating along the domain wall between two parity-transformed lattices (VPC$_1$ and VPC$_2$). We visualize the spatial wavefunction of the mode with a phase-sensitive near-field scanning optical microscope (NSOM) (Fig. \ref{fig:1}b) \cite{Gersen2005a, Rotenberg2014}. Figure \ref{fig:1}c shows the measured two-dimensional in-plane field amplitude map at a wavelength of $\lambda = \SI{1600}{nm}$. The detected transverse-electric (TE)-like field pattern confined to the interface of VPC$_1$ and VPC$_2$ extends laterally over roughly five unit cells, revealing an intricate sub-wavelength mode structure (left inset of Fig. \ref{fig:1}c). The measured fields show close correspondence to numerical calculations (see Supplementary Fig. S1). At the locations of the access and exit waveguide, the influence of broken lattice symmetry and the adjacent feed waveguide becomes evident in the distorted field pattern (right inset of Fig. \ref{fig:1}c).

The heterodyne detection configuration of the employed NSOM gives access to the complex in-plane optical fields of the edge mode \cite{Balistreri2001}. As a direct consequence of Bloch's theorem, the two-dimensional spatial Fourier transformation $\mathcal{F}(k_x , k_y)$ of the measured field amplitude allows individual analysis of Fourier components with positive and negative phase velocities. An illustrative Fourier map at $\lambda = \SI{1600}{nm}$ is displayed in Fig. \ref{fig:2}a. By repeating the near-field scans and corresponding Fourier analysis for $\lambda = [\SI{1480}{nm} - \SI{1640}{nm}]$ and integrating $\mathcal{F}(k_x , k_y)$ over $k_y$, we extract the mode dispersion shown in Fig. \ref{fig:2}b. We resolve at least six parallel lines due to the excellent signal to background ratio (S/B) of $\approx \SI{56}{dB}$. The achieved spatial resolution, combined with the high S/B, enables us to resolve higher-order Bloch harmonics over multiple Brillouin zones. The lines with a positive slope correspond to a single forward-propagating mode with group velocity $v_g = c/6$. Closer inspection reveals negatively-sloped lines corresponding to a single backward propagating mode with $-v_g$ \cite{Engelen2007, Burresi2009a}. This separation of forward- and backward-propagating Bloch modes allows local monitoring of back-scattering along the domain wall.

\begin{figure}[hbt]
  \centering
  \includegraphics[width=0.48\textwidth]{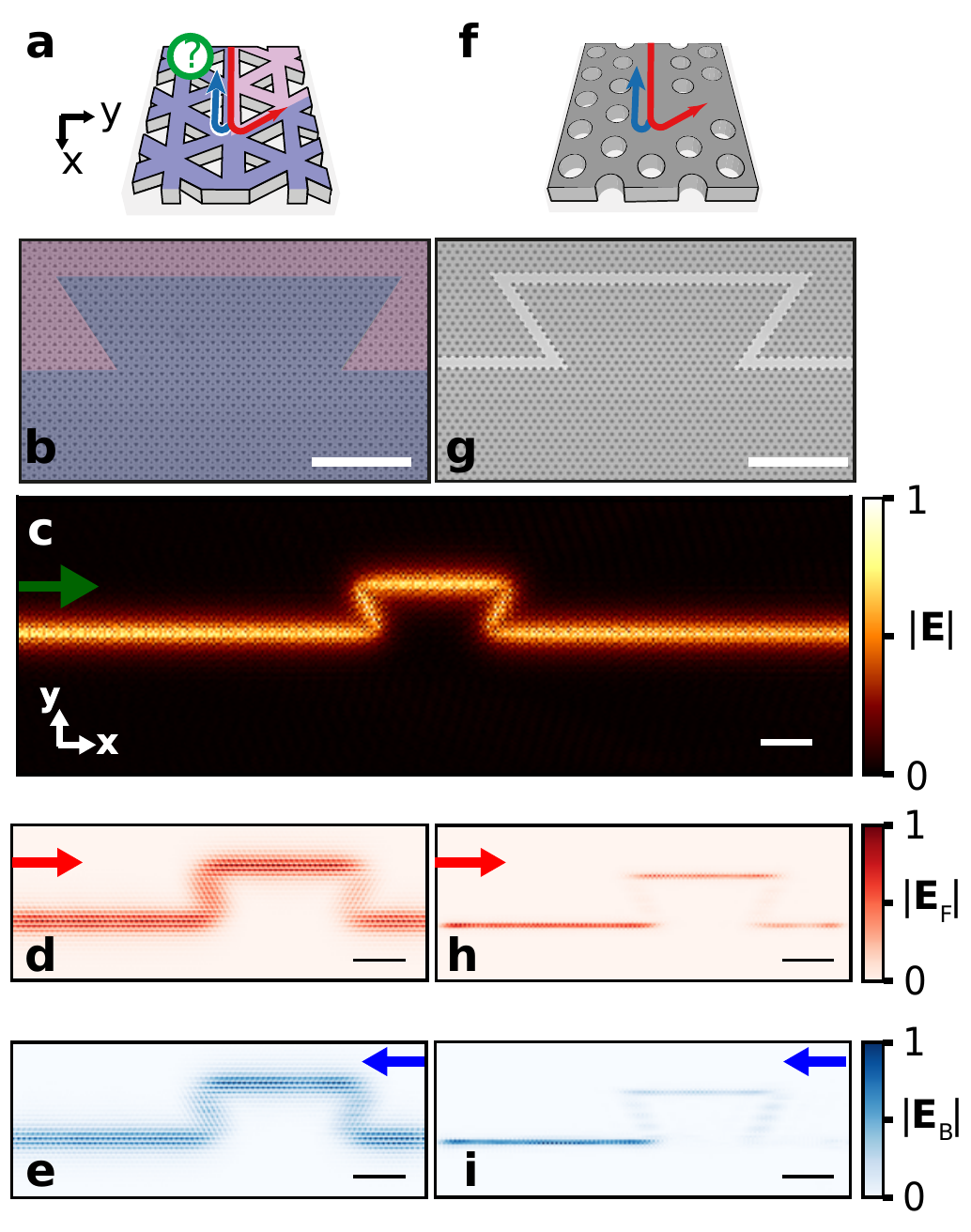}
  \caption{\textbf{Directional transport along defects.} For the topologically non-trivial VPC waveguide, \textbf{a} shows the schematic of the probed $120^{\circ}$ corner.
  \textbf{b}, Top-view SEM image of the fabricated $\Omega$-shaped defect. Two dimensional real-space amplitude maps showing the \textbf{c} full mode amplitude distribution, \textbf{d} forward propagating mode amplitude, and \textbf{e}, backward propagating mode amplitude. The amplitude maps are normalized independently to their maximum value.
  For a topologically trivial W1 waveguide, \textbf{f} shows schematically the mode propagation around a $120^{\circ}$ corner and \textbf{g} a top-view SEM image of the device. The two-dimensional amplitude maps of the forward  and backward  propagating modes are shown in  \textbf{h} and  \textbf{i}, respectively. Scale bar: $\SI{5}{\mu m}$.}
  \label{fig:3}
\end{figure} 

Using this local phase and amplitude information, we probe a straight edge domain wall shown in Fig. \ref{fig:1}c. We obtain the quantities $ W_F $ and $ W_B $ representing the forward and backward energy, respectively, through integration of their corresponding Fourier intensity. The ratio $\eta_e = {W_B}/{W_F} \approx 0.03$ unambiguously yields the conversion from forward to backward propagation, a result of scattering events occurring at and beyond the VPC end facet. Thus, $\eta_e$ includes coupling of the forward to backward mode energy away from the topologically protected regime. This initial examination of the straight edge with the observed back-propagation energy dominated by contributions of the end-facet, calls for a more intricate analysis of topological protection.

\begin{figure}[hbt]
  \centering
  \includegraphics[width=0.48\textwidth]{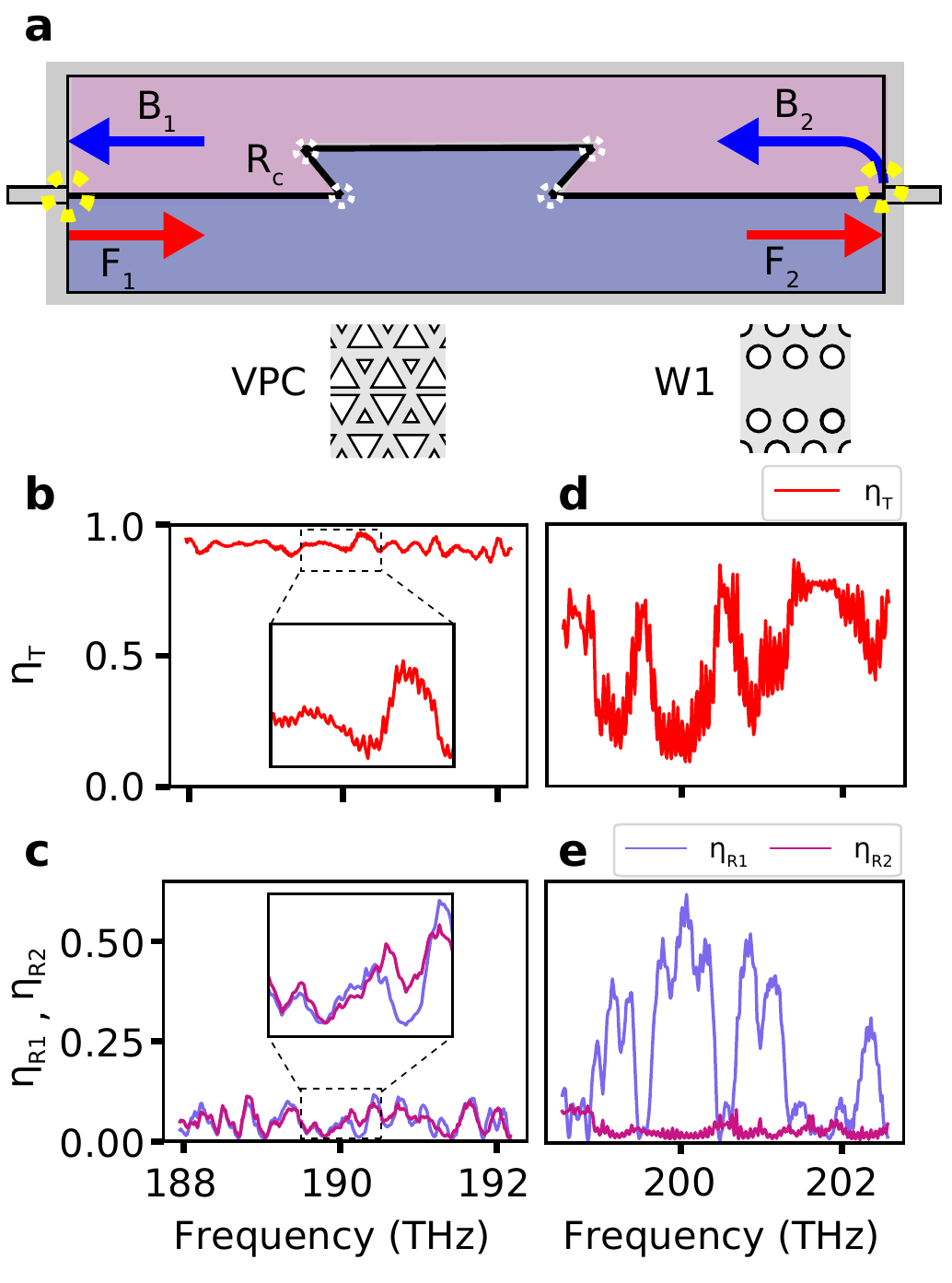}
  \caption{\textbf{Degree of topological protection. a,} Schematic of the mode contributions in a $\Omega$-shaped defect VPC waveguide. Red arrows indicate the forward propagating modes, with $F1$ and $F2$ denoting the modes before and after the defect, respectively. Blue arrows indicate backward propagating modes before  ($B1$) and after ($B2$) the defect. The yellow dashed circles show the locations of in- and outcoupling- facets. White dashed circles indicate the four $\SI{120}{\degree}$ corners. \textbf{b}, Plot of the transmission coefficient $\eta_T$, with the inset demonstrating transmission over a small region([$\SI{189.5}{THz} - \SI{190.5}{THz}$]). \textbf{c} Backward/forward energy ratio before ($\eta_{R1}$) and after ($\eta_{R2}$) the $\Omega$-shaped defect in the VPC domain wall. The inset shows how the back-propagation energies before and after the defect are almost indistinguishable over the considered frequency range. \textbf{d}, and \textbf{e}, show the corresponding plots of \textbf{b} and \textbf{c} for the W1 waveguide, respectively.}
  \label{fig:4}
\end{figure} 

To quantify protection without the aforementioned contributions, we introduce a trapezoidal ($\Omega$-shaped) structure along the domain wall comprising four sharp corners (Fig. \ref{fig:3}). This structure is expected to be topologically-protected since $n \times 120^{\circ}$ bends respect the underlying $C_3$ lattice symmetry. Reflections characterized by energy coupled between the degenerate forward ($F$) and backward ($B$) propagating modes are indicated by red and blue arrows, respectively, in the schematic Fig. \ref{fig:3}a. Figure \ref{fig:3}c shows a map of the measured amplitude of the VPC edge mode. Separating the forward and backward modes through Fourier filtering leads to Fig. \ref{fig:3}d and e. Figure \ref{fig:3}d qualitatively demonstrates that the forward propagating mode exhibits a near-unity transmission through the bend. The constant amplitude of the backward propagating mode (Fig. \ref{fig:3}e) also indicates near-unity transmission. This demonstrates that we may attribute the coupling of the forward and the backward mode to the termination of the exit PhC waveguide. Put differently, light is perfectly guided around the $\Omega$-shaped domain wall, with the transmission being independent of the presence of the defect itself.

This observation is quantified by translating the locally measured amplitudes into mode energy ratios. We filter the Fourier intensity distribution to obtain the forward and backward propagating mode energy before ($W_{F1}, W_{B1}$) and after ($W_{F2}, W_{B2}$) the $\Omega$-bend (see Fig. \ref{fig:4}a). Locally determined transmission through the defect for the linear part of the dispersion is shown in Fig. \ref{fig:4}b. A mean transmission value $\eta_T = {W_{F1}}/{W_{F2}}$ of $ \approx 0.92$ is obtained for the chosen frequency range. Additionally, the mode energy ratios calculated for regions before ($\eta_{R1} = W_{B1}/W_{F1}$) and after ($\eta_{R2} = W_{B2}/W_{F2}$) the defect are shown for a frequency range of $\SI{4}{THz}$ in Fig. \ref{fig:4}c. We notice that $\eta_{R1}$($f$) and $\eta_{R2}$($f$) are almost indistinguishable. This strongly suggests that the contribution of the four symmetry-protected corners to the back-propagation energy is insignificant with respect to back-scattering at the end facet.

Although expected, one can appreciate that the remarkably large transmission over the mode's full frequency range \cite{Ma2015, He2019, Tzuhsuan2016} is reasonably atypical in comparison with a topologically trivial standard ``W1'' PhC waveguide (see Methods for fabrication details). We again introduce a trapezoidal structure in said PhC waveguide (Fig. \ref{fig:3}f, g). It is worth mentioning that the fabricated W1 waveguide corners are not optimised for unity transmission at any given frequency \cite{Krauss2010}.
In stark contrast to the forward and backward mode for a VPC (Fig. \ref{fig:3}d, e), the W1 modes (Fig. \ref{fig:3}h, i) show significant loss across the defect. Moreover, the normalized backward amplitude map in Fig. \ref{fig:3}i demonstrates that the dominant reflections already occur at the first $120^{\circ}$ corner. The mode energy here is converted to a back-reflected wave and additionally experiences out-of-plane scattering loss. The $\eta_T$ measured through the $\Omega$-structure in the W1 PhC, shown in Fig. \ref{fig:4}d, is on average one-third of the $\eta_T$ observed for the VPC. The strong reflection from the first corner is confirmed by $\eta_R$ shown in Fig. \ref{fig:4}e, where $\eta_{R1}$ is four times higher than $\eta_{R2}$ for certain frequencies in the W1 PhC waveguide.

In addition to the back-reflection from the individual corners, the direct evaluation of the $\Omega$-shaped defect is affected by other aspects: out-of-plane scattering losses, scattering at the end-facet and interference due to multiple reflections along the domain wall. We notice rapid oscillations in $\eta_{R1,2} (f)$ before and after the defect (Fig. \ref{fig:4}c,e). To disentangle the back-scattering contribution from the aforementioned effects, we consider the complex scalar mode amplitude of the Bloch wave at different points along the domain wall. With the assumption of a perfectly mirror symmetric device, we treat the defect as a single effective interface in a transfer-matrix model (TMM). Using $\eta_R$ and $\eta_T$ as input parameters to the model, we quantify the mean reflectance $\overline{R_c}$ of the full defect. Details of the model and the precise extraction method can be found in the Supplementary Sec. II. Applying the model to the data for the topologically protected edge states shown in Fig. \ref{fig:4}b,c we retrieve a mean effective reflectance for the full defect $\overline{R_c} = 0.002 \pm 0.001$ and an out-of-plane scattering loss $\overline{A_c} = 0.080 \pm 0.002$ for the topologically protected edge states. Furthermore, we determine the average single-corner reflectance $R_c^{\mathrm{single}} = 0.0007$ from the TMM (see Supplementary Sec.II C).

The same approach applied to the data in Fig. \ref{fig:4}d,e for the W1 PhC waveguide reveals a reflectance of $\overline{R_c} = 0.191 \pm 0.010$, two orders of magnitude larger than that observed for the VPC, and an out-of-plane scattering coefficient $\overline{A_c} = 0.304 \pm 0.017$. These values for the W1 structure are in close agreement with literature \cite{Kuang2004, Chow2001, Chutinan2002}. A topologically protected PhC lattice thus reduces the experimentally achievable back-reflection from individual sharp corners by two orders of magnitude over the entire frequency range of the edge state.

In summary, a direct experimental quantification of topological protection in VPC-based PTIs at telecom frequencies was achieved by accessing the full complex wavefunction of the edge state via phase-resolved near-field microscopy. This allows for determination of back-reflection from topologically protected defects as well as for quantification of the experimentally unavoidable out-of-plane scattering losses. We unambiguously determined an experimental upper limit to the back-scattering contribution from symmetry-protected defects in PhC-based topological edge states. This evaluation opens a direct pathway towards applied quantum topological photonic networks for secure and robust communications. 

\bibliography{QVHE_bibliography}

\section{Methods}

\textbf{Simulations.} Numerical simulations were performed using MIT Photonic Bands \cite{Johnson2001} with the in-plane field distributions and retrieved dispersion relation shown in the supplementary materials. To match the calculated edge state to the measured dispersion relation, the  refractive index of silicon was chosen as $n = 3.36$. In order to account for the corner roundness arising from fabrication, a fillet of $\SI{42}{nm}$ radius was added to the triangular holes of lattice constant $a = \SI{503}{nm}$. The unit cell consisted of equilateral triangles, with a larger triangle side length $d_1 = 0.7 a$ and smaller triangle side length $d_2 = 0.45 a$. 
In addition, finite-difference time-domain calculations (FDTD Solutions by Lumerical) were used to verify the intrinsic transmittance spectra through $120^\circ$ bends in a W1 PhC waveguide.

\textbf{Device fabrication.} The PhC slab was fabricated on a silicon-on-insulator (SOI) platform with a $\SI{220}{nm}$ thick silicon layer on a $\SI{3}{\mu m}$ buried oxide layer. The fabrication was performed in two steps: First, a positive electron-beam resist of thickness $\SI{240}{nm}$ (AR-P 6200.09) was spin-coated between a monolayer of adhesion reagent HMDS and a conductive layer of E-Spacer 300Z. Then, the PhC design was patterned in the resist using e-beam lithography on a Raith Voyager with \SI{50}{kV} beam exposure. The e-beam resist was developed in pentyl acetate/O-Xylene/MIBK:IPA(9:1)/isopropanol, and the SOI chip subsequently underwent reactive-ion etching in HBr and O$_2$. In a second step, the photo-lithography resist AZ1518 was patterned using a Suss MABA6 Mask Aligner to define a selective wet-etching window on the PhC. After development with AZ400K:H$_2$O, the buried-oxide layer was removed in an aqueous 5:1 solution of hydrofluoric acid. The PhC was then subjected to critical point drying before being mounted in the near-field optical microscopy setup. 
The PhCs are terminated on both sides such that a TE single-mode Si-ridge waveguide is extended as PhC waveguide into the crystal to enable better index-matching for efficient incoupling \cite{Shalaev2018}.

The PhC lattice features a honeycomb configuration of two equilateral triangles in a unit cell of lattice constant $a = \SI{503}{nm}$. One triangle is scaled up ($d_1 = 0.7 a$) and the other down ($d_2 = 0.45 a$), while preserving $C_3$ lattice symmetry. A domain wall is created along the VPC$_1$ and VPC$_2$ by applying a parity operation along the spatial $y$-coordinate. Two different VPC domain walls were fabricated to facilitate transmission and back-scattering comparisons. The straight edge VPC has dimensions $195a \times 55a$ designed such that the PBG falls within the tunable laser wavelength range of $1480-\SI{1640}{nm}$. For the trapezoidal edge VPC domain wall, the two diagonals extend over $12$ unit cells, whereas the horizontal extent of the defect between the second and third corner is $34$ unit cells. 

The standard `W1' waveguide is formed from a honeycomb lattice of circular holes, with a lattice constant of $a=\SI{420}{nm}$ and hole radius $r=0.3a$, where one row of circular holes was removed.

\textbf{Near-field optical microscopy setup.} The utilized aperture-based near-field optical microscope consists of a tapered optical fiber coated homogeneously with $\SI{140}{nm}$ aluminium. An aperture of ca. $\SI{170}{nm}$ is created at its apex via focused ion beam milling. Scanning the probe over the silicon membrane at a relative height of $\approx \SI{20}{nm}$ controlled via shear force feedback results in the pickup of the local in-plane field components. Their amplitudes and phases are determined using a heterodyne detection scheme, with the coherent reference light beam shifted by $\Delta f = \SI{40}{kHz}$ in frequency \cite{Rotenberg2014}.

\section{Acknowledgements}
We thank Nikhil Parappurath, Filippo Alpeggiani and Aron Opheij for fruitful discussions about the initial design, fabrication and measurement steps.
This work is part of the research programme of the Netherlands Organisation for Scientific Research (NWO). The authors acknowledge support from the European Research Council (ERC) Advanced Investigator Grant no. 340438-CONSTANS and ERC starting grant no. 759644-TOPP.

\section{Author Contributions}
R.B. fabricated the devices. S.A. and T.B. carried out the near-field measurements. S.A, R.B. and T.B. performed data analysis and modelling. E.V. and L.K. conceived and supervised the project. All authors contributed extensively to the interpretation of results and writing of the manuscript.

\section{Data availability}
All data obtained in the study are available from the corresponding author upon reasonable request.

\end{document}


\title{Supplementary information: Direct quantification of robust topological protection in photonic crystals supporting symmetry-protected edge states at telecom wavelengths}

\author{S. Arora}
\affiliation{Kavli Institute of Nanoscience, Delft University of Technology, 2600 GA, Delft, The Netherlands}
\author{T. Bauer}
\affiliation{Kavli Institute of Nanoscience, Delft University of Technology, 2600 GA, Delft, The Netherlands}
\author{R. Barczyk}
\affiliation{Center for Nanophotonics, AMOLF, Science Park 104, 1098 XG Amsterdam, The Netherlands}
\author{E. Verhagen}
\affiliation{Center for Nanophotonics, AMOLF, Science Park 104, 1098 XG Amsterdam, The Netherlands}
\author{L. Kuipers}
\affiliation{Kavli Institute of Nanoscience, Delft University of Technology, 2600 GA, Delft, The Netherlands}

\maketitle
\onecolumngrid

\section*{Glossary}
\noindent
VPC --- valley photonic crystal\\
TMM --- transfer matrix model\\
\section{Numerical calculation of edge state dispersion and eigenmodes}
The edge state eigenfrequencies $f(k)$ as well as eigenmodes $\mathbf{E}_{k}(\mathbf{r})$ that are localized at the domain wall between two mirror-symmetric valley photonic crystals (VPCs) were calculated for an in-plane wavevector $k$ with the freely available MIT Photonic Bands solver\,\cite{Johnson2001}. It determines the Bloch eigenmodes of the full three dimensional photonic crystal structure using a plane-wave basis set and periodic boundary conditions. We used a simulation supercell of dimensions $a \times 28a \times 10h$, where $h$ is the thickness of the free-standing silicon slab and $a$ is the lattice constant of the VPC patterned therein. This supercell was sufficiently large to avoid interactions between neighbouring supercells in $y$ and $z$. The calculations used an in-plane grid size of $a/32$ and out-of-plane grid size of $a/16$, which ensured convergence of the eigenvalues to better than $0.1\%$. To account for surface passivation effects of the experimentally employed silicon membrane of height $h=\SI{220}{nm}$ and to adapt the resulting eigenfrequencies to the experimentally determined dispersion relation, the effective refractive index of silicon was modelled to be 3.36.

\begin{figure}[hbt]
  \centering
  \includegraphics[width=\textwidth]{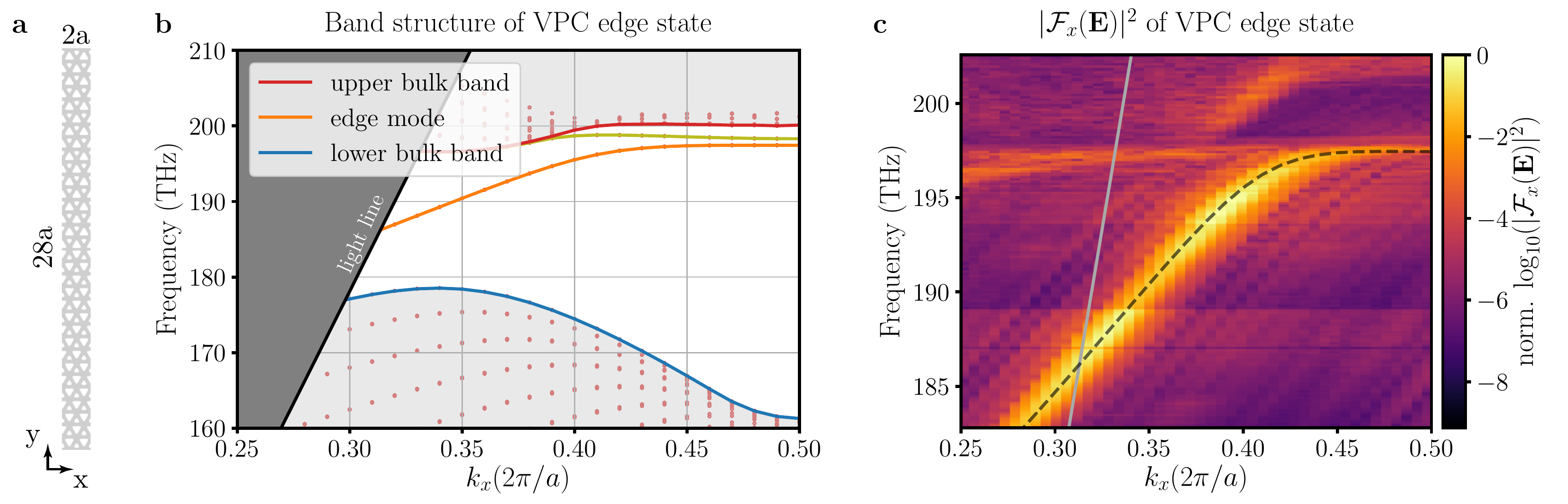}
  \caption{Band structure calculations. \textbf{a} Geometry of the photonic crystal membrane used in the eigenmode calculations, with grey and white representing silicon and air, respectively. For better visibility, two unit cells are shown along the domain wall. \textbf{b} Numerically calculated band structure of the edge state eigenmode (orange) as well as the upper and lower bulk bands (red and blue, respectively) of the photonic crystal. The light cone is displayed as dark grey shading, while the bulk band region has been shaded light gray. In addition to the topologically non-trivial edge mode, there exists a second mode confined to the edge at large $k_x$, shown in ochre. \textbf{c} Comparison of the fundamental Brillouin zone of the experimentally retrieved dispersion relation and the numerically calculated edge state eigenfrequencies (black dashed line), highlighting the excellent overlap between experiment and simulation. The light line is given by the solid grey line.}
  \label{fig:supp1}
\end{figure}
The lattice constant was chosen as $a=\SI{503}{nm}$, and the base length of the triangular holes in each unit cell,  determined from SEM images, was $d_1=0.70a$ and $d_2=0.45a$ for the large and small triangle, respectively. Both triangular holes were modeled with a corner rounding of $r=\SI{42}{nm}$ to account for fabrication-related deviations from their ideal structure. We construct a bridge-type domain wall between two mirror-symmetric sub-lattices VPC$_1$ and VPC$_2$ in the supercell, with the base of the large triangles facing each other. The periodic nature of the simulation subsequently dictates the additional occurrence of the complementary domain wall (base of small triangles facing each other) at the boundary of the supercell (see Fig. \ref{fig:supp1}\textbf{a}). The calculated edge state eigenmodes located at this complementary domain wall were filtered out, and the resulting band structure for the topologically protected edge state as well as the bulk VPC eigenfrequencies determining the system's band gap are shown in Figure \ref{fig:supp1}\textbf{b}. In addition to the edge state traversing the band gap with a linear dispersion around the K-point, a second mode confined to the central domain wall is found branching off the top bulk band for high in-plane $k$-values (shown in ochre in Fig. \ref{fig:supp1}\textbf{b}). Comparing the calculated band structure to the fundamental Brillouin zone of the experimentally retrieved dispersion relation, an excellent agreement in both position of the band edge as well as edge state dispersion is found, with the mode dispersion of the edge state extending into the light cone for $k_x < 0.31 \cdot {2 \pi}/{a}$ (see Fig. \ref{fig:supp1}\textbf{c}).

\begin{figure}[hbt]
  \centering
  \includegraphics[width=\textwidth]{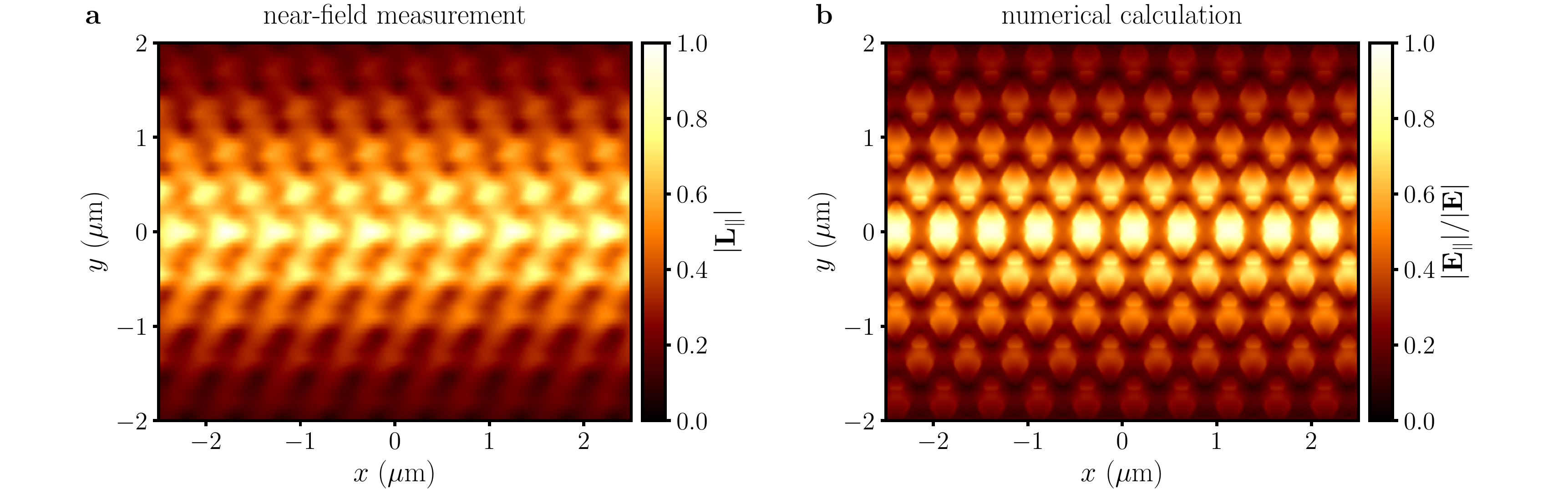}
  \caption{Comparison between \textbf{a} the experimentally retrieved near-field amplitude map for $\lambda = \SI{1600}{nm}$ ($f = \SI{187.4}{THz}$) and \textbf{b} the numerically calculated in-plane electric field distribution of the edge mode at $k_x / (2 \pi /a)=0.32$ ($f = \SI{187.0}{THz}$).}
  \label{fig:supp2}
\end{figure}

To verify the experimentally determined near-field structure of the edge mode (shown for a wavelength of $\lambda = \SI{1600}{nm}$ in Fig. \ref{fig:supp2}\textbf{a}), the field distribution of the eigenstate was extracted from the simulation $\SI{20}{nm}$ above the silicon membrane. The in-plane field amplitude distribution is displayed for an illustrative $k$-point of $k_x=0.32 \cdot {2 \pi}/{a}$ in Fig. \ref{fig:supp2}\textbf{b}. The corresponding eigenfrequency of the edge state at this $k$-point is $f=\SI{187.0}{THz}$. It can be seen that the mode symmetries of this numerically determined eigenmode match for the mode's in-plane electric field amplitude and the experimentally measured mode distribution. 

\section{Quantification of corner reflectivity and loss}\label{ch:quan}
Our main goal in this section is to quantify the back-scattering of topologically protected edge states -- bound to a VPC domain wall -- from an $\Omega$-shaped defect of four $120^{\circ}$ corners (cf. Fig. 3\textbf{b} of the main manuscript) by comparing the measured forward and backward propagating energies before and after the defect to an analytical model. We introduce a transfer matrix model (TMM) approach and compare its results to the back-scattering observed in a similarly shaped W1-type photonic crystal waveguide, which is based on one missing row of holes in a hexagonal lattice. We extract the  reflection coefficient $R_c^\text{single}$ of an individual corner in the waveguide as a quantitative measure of back-scattering, and quantify the associated loss $A_c^\text{single}$ which is dominated by out-of-plane scattering.
\subsection{Transfer matrix model}\label{ch:tmm}
\begin{figure*}[!htb]
\centering
\includegraphics[width=\textwidth]{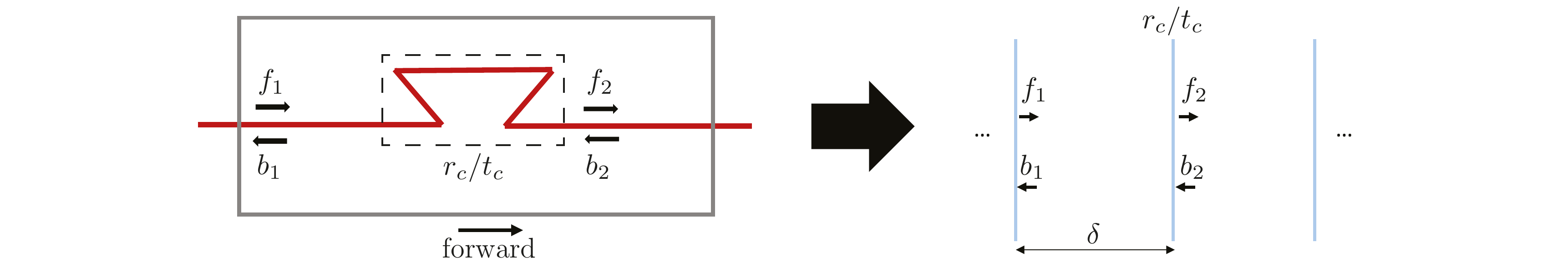}
\caption{Sketch of the experimentally investigated VPC with an $\Omega$-shaped domain wall defect, as well as its representation within a transfer matrix model. The effect of the $\Omega$-shaped defect onto the field amplitudes is approximated by complex reflection and transmission coefficients $r_c$ and $t_c$, and the distance between the detection points is associated with a propagation phase $\delta$.}
\label{fig:tmm_scheme}
\end{figure*}

We start from Bloch's theorem for a photonic crystal domain wall (or waveguide) with periodicity $a$ along the $x$-direction, resulting in the Bloch state 
\begin{align}
\mathbf{E}_{k_x}(\mathbf{r})= e^{\imath {k_x x}} \cdot \mathbf{u}_{k_x}(\mathbf{r}) \, , \quad  \text{with } \mathbf{u}_{k_x}(\mathbf{r} + a \hat{\mathbf{e}}_x) = \mathbf{u}_{k_x}(\mathbf{r}),
\end{align}
where $k_x$ is the Bloch wavevector along the mode's propagation direction, and $\mathbf{u}_{k_x}(\mathbf{r})$ represents its full spatial field distribution.

To compare the mode at different positions along the domain wall, we separate a complex scaling amplitude $a_k$ from the mode's periodic distribution, resulting in
\begin{align}
\mathbf{E}_{k_x}(\mathbf{r})= a_k e^{\imath {k_x x}} \cdot \mathbf{u}{'}_{k_x}(\mathbf{r}),
\end{align}
with $\mathbf{u}{'}_{k_x}$ a fixed function, independent of the unit cell or time at which the mode field is evaluated. The full spatial information in each unit cell is thus effectively reduced to the scalar complex amplitude value $a_k$.

The complex amplitude together with the reciprocal nature of the employed photonic crystals enable us to describe back-scattering sites along the domain wall as interfaces with complex amplitude reflection and transmission coefficients $r_c$ and $t_c$, respectively. 
Furthermore, we introduce a loss channel $A_c$, incorporating out-of-plane scattering losses, via
\begin{align}
R_c+T_c = 1-A_c \,, \label{eq:loss}
\end{align}
where  $R_c = |r_c|^2$, $T_c = |t_c|^2$.

Without loss of generality, we consider the forward and backward propagating field amplitudes $f_1, f_2$ and $b_1, b_2$ to denote the value of $a_k$ at specific positions on the sample as shown in \figref{fig:tmm_scheme}, and get
\begin{align}
b_1 &= r_c  f_1\, e^{i2\delta} + t_c b_2\, e^{i\delta} \label{eq:amp1}  \\
f_2 &= t_c f_1\, e^{i\delta}  + r_c b_2  \, , \label{eq:amp2}
\end{align}
with the propagation phase $\delta=k\cdot d$ given by the wavenumber $k$ and distance $d$ between the two detection points 1 and 2, which are placed immediately after an interface. Equations \ref{eq:amp1}, \ref{eq:amp2} can be cast into the compact form
\begin{align}
  \begin{pmatrix}
    f_1  \\
    b_1 
  \end{pmatrix}
= \mathbf{M}_c  
\begin{pmatrix}
    f_{2}  \\
    b_{2} 
  \end{pmatrix} \,,
\end{align}
where
\begin{align}
\mathbf{M}_c =
\frac{1}{t_c}
\begin{pmatrix}
    e^{-i\delta} & 0  \\
    0 &  e^{i\delta}
\end{pmatrix}
\begin{pmatrix}
    1 & -r_c  \\
    r_c &  t_c^2 - r_c^2
\end{pmatrix} 
\end{align}
is the transfer matrix describing light propagation from position 1 to 2 in the multilayer stack.The propagation through a series of $N$ layers can then be written as a single matrix $\tilde{M}$, given by
\begin{align}
\mathbf{\tilde{M}} = \sum_{i=1}^N \mathbf{M}_i \,.
\end{align}
Here, $\mathbf{M}_i $ are the transfer matrices of the individual layers (with corresponding distances and reflection/transmission coefficients).

\begin{figure*}[!htb]
\centering
\includegraphics[width=\textwidth]{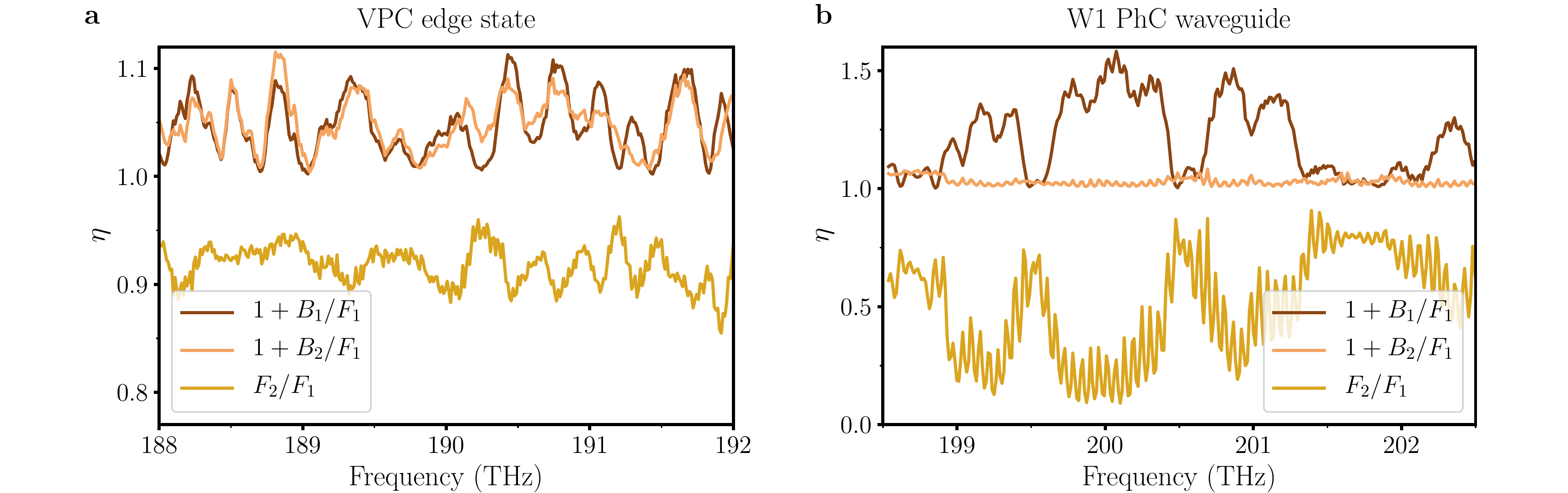}
\caption{Experimentally retrieved power of the forward ($F_2$) and backward ($B_1, B_2$) propagating modes before and after the $\Omega$-shaped defect, normalized to $F_1$. The backward mode intensities are shifted by 1 for better visibility. \textbf{a} VPC edge state. \textbf{b} W1 waveguide.}
\label{fig:normF1}
\end{figure*}

It is straightforward to relate any field amplitudes in a TMM to each other by appropriate transfer matrix multiplications. The experimentally determined power densities on which we base our analysis are given by $F_{n}=|f_{n}|^2$, $B_{n}=|b_{n}|^2 \ (n=1,2)$ and are repeated from Fig. 4\textbf{b}-\textbf{e} of the main manuscript in \figref{fig:normF1} over an excitation frequency window of ca. $\SI{4}{THz}$.  If we want to achieve a full overlap of these power densities with predictions from the TMM model, we not only need to take into account the four $120^\circ$ corners of the tailored topologically protected defect, but additionally all scattering sites after the defect. These include scattering at the transition from the VPC domain wall to an intermediate PhC waveguide, the subsequent transition to a silicon strip waveguide, and scattering at potential fabrication defects. Since these additional interfaces are not protected by the mirror symmetry of the VPC lattice, it is expected that their reflection coefficients dominate the entire system.

\subsection{Effective corner reflectivity}\label{ch:eff}
\begin{figure*}[!htb]
\includegraphics[width=\textwidth]{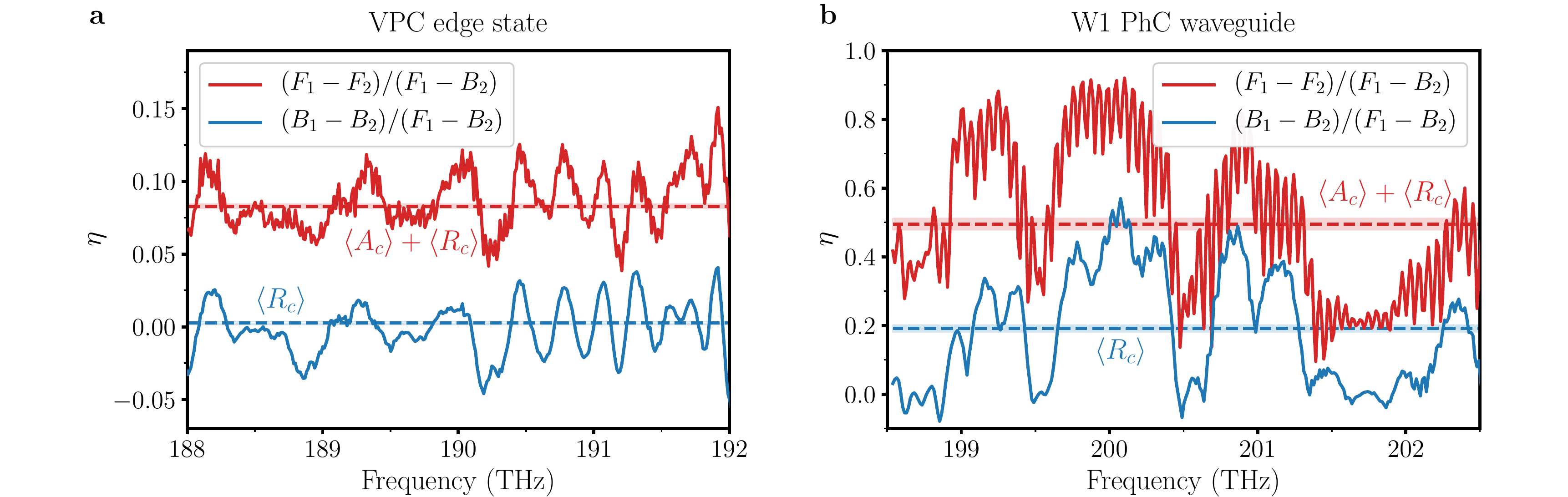}
\caption{Evaluation of \eqref{eq:power_loss_sym1},~\eqref{eq:power_loss_sym2} for the data shown in Fig. \ref{fig:normF1}, where $\left\langle R_c\right\rangle$ and $\left\langle R_c\right\rangle + \left\langle A_c \right\rangle$ according to \eqref{eq:Rc_mean},~\eqref{eq:Ac_mean} are indicated as red and blue dashed lines, respectively. The shaded region around the lines corresponds to the standard deviation of the extracted mean values. \textbf{a} VPC edge state. \textbf{b} W1 waveguide.}
\label{fig:Rc_mean}
\end{figure*}

Our aim is to arrive at a relation between the measured power densities before and after the defect, and the intensity reflection coefficient $R_c$ and loss $A_c$. If we, in fair approximation, assume our devices to be mirror symmetric (as by design), reciprocity dictates equal coefficients for forward and backward propagation through the full $\Omega$-shaped defect. We can thus combine the four individual corners into a single effective interface with complex amplitude reflection and transmission coefficients $r_c$ and $t_c$ (see \figref{fig:tmm_scheme}).
We divide \eqref{eq:amp1},~\eqref{eq:amp2} by $f_1$ and take the absolute square to obtain
\begin{align}
\frac{B_1}{F_1} &= R_c + \frac{B_2}{F_1}\, T_c + \left(\sqrt{R_c} e^{-i\theta_{r_c}}\, \sqrt{T_c} e^{i\theta_{t_c}}\, \sqrt{\frac{B_2}{F_1}}  e^{i\theta_{b_2}}  e^{-i\theta_{f_1}} + c.c.\right) \nonumber\\ &= R_c + \frac{B_2}{F_1}\, T_c + 2\sqrt{R_c\, T_c\, B_2\, F_1} \cos(\theta_c-\theta) \label{eq:amprat1}\\
\frac{F_2}{F_1} &= T_c + \frac{B_2}{F_1}\, R_c + 2\sqrt{R_c\, T_c\, \frac{B_2}{F_1}}\ \ \ \cos(\theta_c+\theta) \,. \label{eq:amprat2}
\end{align}
Here,  $c.c.$ denotes the complex conjugate and we express complex quantities in their polar form ($z=\sqrt{Z} e^{i\theta_z}$), with the phase differences $\theta_c = \theta_{r_c}-\theta_{t_c}$ and $\theta = \theta_{b_2}-\theta_{f_1}$. Using \eqref{eq:loss}, we cast \eqref{eq:amprat1},~\eqref{eq:amprat2} into the form
\begin{align}
\frac{B_1-B_2}{F_1 - B_2} = R_c - \frac{A_c B_2 - 2\sqrt{R_c(1-R_c)F_1 B_2}}{F_1 - B_2} \, \cos(\theta_c-\theta)  \label{eq:power_loss_sym1} \\
\frac{F_1-F_2}{F_1 - B_2} = R_c + \frac{A_c F_1 - 2\sqrt{R_c(1-R_c)F_1 B_2}}{F_1 - B_2} \, \cos(\theta_c+\theta) \label{eq:power_loss_sym2}
\end{align}
Equations \eqref{eq:power_loss_sym1}, \eqref{eq:power_loss_sym2} represent the central result of the TMM analysis, and the corresponding data is presented in \figref{fig:Rc_mean}.\\

Due to the expected strong influence of scattering events from outside the topological photonic crystal, as seen by the strong correlation between the signals $B_1/F_1$ and $B_2/F_1$ in Fig. \ref{fig:normF1}\textbf{a}, we utilize an effective averaging approach. Such an approach is based on the linear slope of the VPC edge state dispersion. We can then disentangle the reflection at the tailored $\Omega$-shaped defect from any subsequent back-scattering events by averaging over a frequency range where the interference effects between the latter lead to cosine-like intensity modulations with multiple full periods. Both $\theta_c$ and $\theta$ in \eqref{eq:power_loss_sym1},~\eqref{eq:power_loss_sym2} are oscillatory functions of frequency since they incorporate the propagation phases (cf. Eqs. \ref{eq:amp1}, \ref{eq:amp2}). Therefore, by applying a frequency average over the experimental range of ca. $\SI{4}{THz}$ (denoted by $\left\langle \mathord{\cdot} \right\rangle$) to \eqref{eq:power_loss_sym1},~\eqref{eq:power_loss_sym2}, we average out the cosine modulation 
and solve for the mean effective corner reflectivity and loss
\begin{align}
\left\langle R_c \right\rangle &= \frac{(\,\overline{B_1-B_2}\,)\overline{F}_{1}-(\,\overline{F_1-F_2}\,)\overline{B_2}}{\overline{F_1}+\overline{B_2}} \label{eq:Rc_mean}\\
\left\langle A_c \right\rangle &= \frac{(\,\overline{F_1-F_2}\,)-(\,\overline{B_1-B_2}\,)}{\overline{F_1}-\overline{B_2}} \label{eq:Ac_mean} \,,
\end{align}
where we defined the operator
\begin{align}
\overline{\mathord{\cdot}} \equiv \left\langle \frac{\mathord{\cdot}}{F_1 - B_2} \right\rangle \,,
\end{align}
and assume no correlation in the spectral behavior of $A_c$ and $F_1$, $B_2$. The resulting averages are displayed as straight lines in \figref{fig:Rc_mean}, with their standard deviation indicated by the shaded regions.

\subsection{Single corner reflectivity}\label{ch:sing}
From the mean effective interface values $\left\langle R_c \right\rangle$ and $\left\langle A_c \right\rangle$ via \eqref{eq:Rc_mean},~\eqref{eq:Ac_mean}, we infer an estimate for the coefficients of the individual corners, i.e., when representing each of the four waveguide bends by a single interface in the TMM. The distances between the interfaces are set to the design parameters, as indicated in \figref{fig:tmm_scheme}. We model a waveguide with four identical corners, then search for an $\left\langle R_c^\text{single} \right\rangle$ such that the mean effective $\left\langle R_c \right\rangle$ given by \eqref{eq:Rc_mean} reproduces the value extracted from the experiment. 
The single corner absorption is subsequently given by
\begin{align}
\left\langle A_c^\text{single} \right\rangle = 1-\left\langle R_c^\text{single} \right\rangle-\sqrt[4]{1-\left\langle R_c \right\rangle-\left\langle A_c \right\rangle} \,.
\end{align}
The mean effective parameters extracted from our measurements of the VPC domain wall and W1 waveguide are displayed in \figref{fig:Rc_mean} and listed together with the single corner parameters in Tab.\,\ref{tab:Rc_A}.
For the VPC edge state defect, we obtain $\left\langle R_c^\text{single} \right\rangle\approx 0.1\%$ and $\left\langle A_c^\text{single} \right\rangle\approx 2\%$. While the value of $\left\langle A_c^\text{single} \right\rangle$ might seem large for a topological photonic insulator, topological protection is only given for in-plane propagation, with out-of-plane confinement in the silicon slab solely provided by total internal reflection. 
If we apply the described model to the experimentally measured mode amplitudes of a topologically trivial W1 photonic crystal waveguide, we obtain a corner reflectivity that is two orders of magnitude higher with respect to the topologically protected system ($\left\langle R_c^\text{single} \right\rangle \approx 5\%$), while we also see roughly a four-fold increase in out-of-plane scattering ($\left\langle A_c^\text{single} \right\rangle \approx 11\%$). Such an increase in losses matches well with the expected  scaling behaviour with the modes' group velocity of $v_g
^{-2}$ \cite{Engelen2008}, with $v_g \approx c / 5.8$ for the VPC edge state and $v_g \approx c / 10.1$ for the W1 waveguide mode.

\begin{table*}[tb]
\centering
\begin{tabular}{ l|c|c } 
  & VPC edge state & W1 waveguide \\ 
 \hline
 $\left\langle R_c \right\rangle$ & $0.0026 \pm 0.0010$ &  $0.191 \pm 0.010$ \\ 
 $\left\langle A_c \right\rangle$ &  $0.0804 \pm 0.0014$ & $0.304 \pm 0.017$ \\ 
  $\left\langle R_c^\text{single} \right\rangle$ &  $0.0007$ & $0.052$\\ 
 $\left\langle A_c^\text{single} \right\rangle$ &  $0.0208$ & $0.105$\\ 
\end{tabular}
 \caption{Comparison of extracted effective mean corner reflectivity and loss for the $\Omega$-shaped defect within the studied photonic crystal waveguides, as well as the estimates for single corners.}
 \label{tab:Rc_A}
\end{table*}

In consequence, the extracted reduction of backreflection between the two waveguide systems highlights the inherent advantage of topological photonic systems even in experimentally achievable conditions.

\section{Dispersion relation of W1 photonic crystal waveguide}
\label{supp_sect:W1dispersion}

\begin{figure}[ht]
      \centering
      \includegraphics[width=\textwidth]{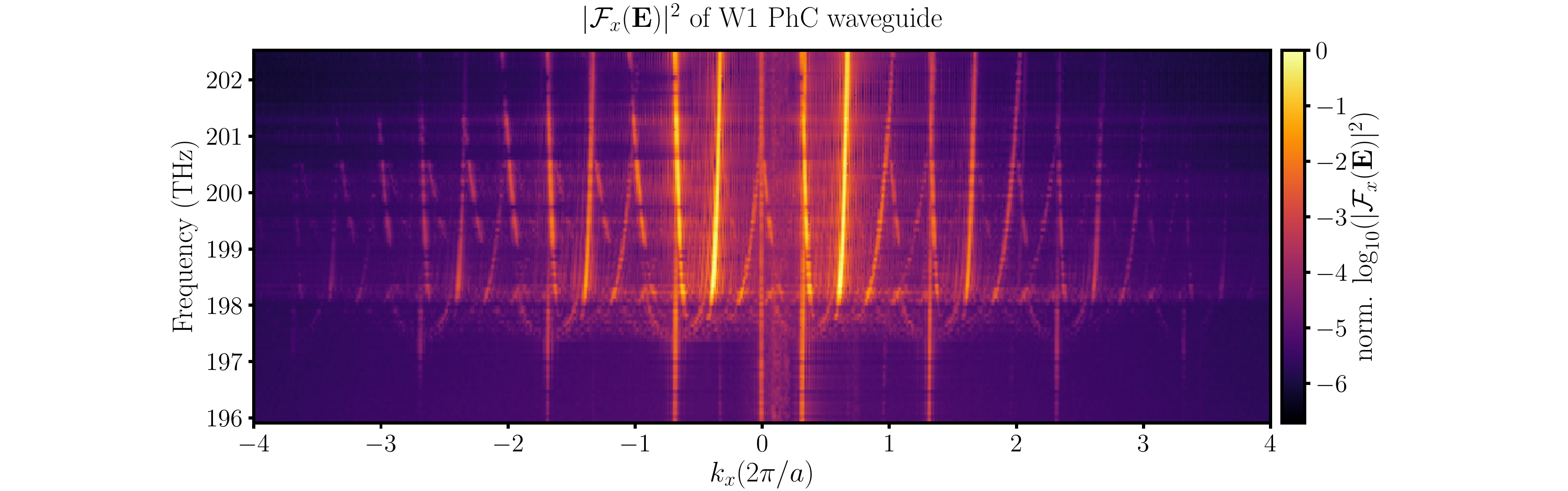}
      \caption{Dispersion relation of the fundamental mode of a W1-type photonic crystal waveguide, retrieved via Fourier transform of near-field maps of the mode while scanning the excitation wavelength. The lattice constant of the investigated photonic crystal is $a=\SI{420}{nm}$.}
      \label{fig:suppl_dispersionW1}
    \end{figure}

To compare the robustness against reflection from a corner defect between the VPC edge state and a W1-type photonic crystal waveguide, both modes need to be considered at comparable group velocities $v_g$, since loss scales with $v_g^{-2}$ \cite{Engelen2008}. Thus, we experimentally determine the dispersion relation of the W1-type photonic crystal waveguide (analogous to the VPC edge state dispersion from the main manuscript, shown in Fig. 2\textbf{b}), and show the bands of forward and backward propagating Bloch modes spanning several Brillouin zones (BZs) in Fig. \ref{fig:suppl_dispersionW1}. One key observation here is that the forward propagating mode in the fundamental BZ is located at negative $k_x$, in accordance with the mode's expected behaviour \cite{Engelen2008}. 
In addition to the predicted band in forward and backward direction, bands with half and a third of the dispersion slope of the fundamental mode appear. These were already observed in a GaAs-based photonic crystal and attributed to a nonlinear interaction with the scanning near-field probe \cite{Singh2015}.

\bibliography{supp_qvhe}